\documentstyle[aps,epsf,epsfig]{revtex}
\begin{document}
\draft
%\twocolumn                 

\title{Probing Vortex Unbinding via Dipole Fluctuations}

\author{H.A. Fertig and Joseph P. Straley}
\address{Department of Physics and Astronomy,
University of Kentucky, Lexington, KY 40506-0055}

%\date{\today}

\maketitle

\begin{abstract}
We develop a numerical method for detecting a vortex
unbinding transition in a two-dimensional system
by measuring large scale fluctuations in the total vortex
dipole moment ${\vec P}$ of the system.  These are characterized
by a quantity $\cal F$ which measures the number of configurations
in a simulation for which the either $P_x$ or $P_y$ is
half the system size.  It is shown that $\cal F$ tends to a non-vanishing
constant for large system sizes in the unbound phase,
and vanishes in the bound phase.  The method is applied
to the $XY$ model both in the absence and presence of
a magnetic field.  In the latter case, the system size dependence
of $\cal F$ suggests that there exist three distinct phases, one unbound
vortex phase, a logarithmically bound phase, 
and a linearly bound phase.
\end{abstract}


\vspace{0.5truecm}
{\it Introduction --}
Topological defects play a crucial role in two dimensional
classical systems\cite{nelson} and in 1+1 dimensional
quantum systems\cite{gogolin}.  The paradigm of these
is the $XY$ model, which is known
to undergo a vortex-unbinding transition in the Kosterlitz-Thouless
(KT) universality class \cite{kosterlitz}.  Highly analogous
transitions occur for vortices in superfluids and thin-film
superconductors, as well as for dislocations and 
disclinations in two dimensional crystals.  Wordline dislocations
in one dimensional quantum systems are important
for describing tunneling \cite{straley}, and the question
of whether these are bound or unbound is closely related
to whether the system is metallic or insulating.

The KT phenomenology has been highly successful in
describing defect unbinding in a variety of situations.
For the $XY$ model, fitting to the expected finite-size
scaling of the helicity modulus in simulations
yields an
estimate of $T_{KT}=0.892$ \cite{olsson}.
However, the concepts of vortices and
defect unbinding are more general than the systems
in which a KT transition takes place.
In some situations a symmetry-breaking field, 
for example, a magnetic field tending
to align the spins in the $XY$ model,  
may be present \cite{jose}.
A recent renormalization group (RG) 
study \cite{decon} has suggested that 
vortex unbinding
does occur in such systems, although the
transition is considerably altered from the
KT behavior.  It is thus important to develop
criteria by which one may determine
whether a given system is in a bound or unbound
vortex state which are independent of precise
matching to the KT theory.  In this article, 
we present a method by which this may be
accomplished, which focuses on a measure
$\cal F$ of extreme
fluctuations in the system vortex dipole moment.
Using a Langevin dynamics simulation of the $XY$ model,
we show that the method locates 
$T_{KT}$ with reasonable accuracy.
We then include a magnetic field
in the Hamiltonian, and show that there
are both bound and unbound vortex phases.
The bound phase has two
distinct behaviors: for smaller fields,
$\cal F$ vanishes as a power of the system size $L$;
for larger fields, $\cal F$ vanishes exponentially with
$L$.  The two behaviors are consistent with the
results of the RG analysis \cite{decon}, which
predicted both a logarithmically bound vortex
phase and a linearly confined one, in addition
to the unbound (deconfined) vortex phase.

{\it Characterizing Vortex Dipole Fluctuations --}
 A total vortex dipole moment may be defined for configurations
of
any system containing vortices with well-defined locations.
For concreteness, we will work with the $XY$ model (planar
spins of fixed length) on a square lattice.  Let 
$\Delta\theta_{ij}$ denote the angular difference between
nearest neighbor spins $ij$, which we reduce to 
the interval $-\pi < \Delta\theta_{ij} \le \pi$
by adding or subtracting $2 \pi$ if necessary.
The vorticity $q_i$ around an elementary plaquette $P$
is then $q_i= (1/2 \pi ) \sum_P \Delta \theta_{ij}$,
which takes on the values $-1,$ $0$, $1$\cite{huh?}.
For any configuration of the $XY$ system
this rule assigns vortex charges to the 
sites ${\vec R}_{i}$ of the dual lattice.
The corresponding dipole moment could be defined by
\begin{equation}
\label{dipo}
{\vec P} = \sum q_i {\vec R}_i  .
\end{equation}
However, there is a problem associated with the use of periodic
boundary conditions:
for a
$L \times L$ system one may add 
$(n^x_i L,n^y_i L)$ to ${\vec R}_i$ (with $n^x_i,~n^y_i$ integers)
and retain a perfectly sensible definition of ${\vec P} $.
For a given configuration, we can use this to reduce $\vec P$ 
so that its components are restricted to the interval
$-L/2 \le P_{x,y} \le L/2$; or we can 
extend its definition by
adding factors of $L$
so that $\vec P$ remains a continuous
function as a vortex crosses a periodicity boundary.
We will refer to these alternate definitions of the dipole moment
as ${\vec P}_{red}$ and ${\vec P}_{ext}$, respectively.
Note that $P^{x,y}_{red}$ will jump discontinuously by $L$
whenever a vortex crosses a boundary.

The bound vortex phase has the property that the dipole moment remains
finite, since the pairs do not separate; in the unbound phase, the 
diffusion of vortices implies a diffusion of the dipole moment, so that
its magnitude can become arbitrarily large.  In a finite system,
this presumably translates into the statement that the diffusion of
 ${\vec P}_{ext}$ becomes very small below the unbinding
temperature.
 
A feature of vortex unbinding transitions is that the transition
occurs via rare but extreme fluctuations \cite{antunes}.
One way to characterize such fluctuations is to look for
system configurations with extreme values of ${\vec P}_{red}$.
Loosely speaking, if we wish to characterize a configuration
in terms of an effective single vortex-antivortex pair, when $P_x$ or $P_y$
is $L/2$, the pair is at its maximum separation.  If such
extreme configurations persist as $L \rightarrow \infty$, 
the system is then in an unbound phase.  By contrast,
we expect the number of such configurations to vanish
in the large size limit when the system is in a bound phase.

This expectation may be more carefully justified by considering
effective theories of vortices in their bound and unbound
phases.  An unbound vortex system behaves as a
two-dimensional metal in which the discreteness of the underlying
vortex charges may be neglected.  An effective Hamiltonian
takes the form \cite{foltin}
\begin{equation}
\label{continuum}
H_{unb}^{eff} = {K \over 2} \int d^2x d^2x'
\rho({\vec x}) G(|{\vec x}-{\vec x'}|) \rho({\vec x'})
+ {{\mu} \over 2} \int d^2x \rho^2({\vec x}),
\end{equation}
where
$G(r) \sim -(2\pi)^{-1} \log(r/a) +$const, with $a$ a microscopic
length of order the lattice constant of the system, and $\mu$ is an
effective chemical potential, essentially the core energy
of the vortices.  For this continuum system, we can adopt
a definition of the ($x$-component of the) dipole moment which incorporates
the periodic boundary condition,
\begin{eqnarray}
P^{x}_{eff} &= {L \over {2\pi}} \int d^2x \rho({\vec x}) \sin \bigl[ {{2\pi x} \over L} \bigl]
\nonumber \\
&= {L \over {2\pi}} Im \lbrace \rho(k_x=2\pi/L,k_y=0) \rbrace .
\label{peff}
\end{eqnarray}
The fraction ${\cal F}$ of configurations with $P^{x}_{eff} = L/2$ is
thus given by the probability of finding $Im \lbrace \rho(k_x=2\pi/L,k_y=0)=\pi$.
This is easily evaluated by re-expressing $H_{unb}^{eff}$ in
terms of a wavevector sum instead of a real space integral.
Noting that the Fourier transform of $G(r)$ 
for small wavevectors is $G(k) \sim 1/k^2 $, one 
obtains
\begin{equation}
{\cal F} \propto e^{-{1 \over {L^{2}T}} [ {1 \over 2} K \bigl( {L \over {2\pi}} \bigr)^2 + \mu ]}
\approx e^{-K/8\pi^{2}T}
\end{equation}
for the probability of obtaining an extreme dipole fluctuation in the large
size limit.  Notice the ${\cal F}$ does not vanish as $L \rightarrow \infty$,
supporting our argument above that large dipole fluctuations survive in
the unbound vortex state.
 
To analyze the bound vortex state, we focus on the two dimensional
Coulomb particle Hamiltonian
\begin{equation}
H_{CG}= {K \over 2} \int d^2x d^2x'
m({\vec x}) G(|{\vec x}-{\vec x'}|) m({\vec x'}),
\end{equation}
where $m({\vec x})$ is an integer degree of freedom, and
the partition function involves a sum over all complexions
of $m$ satisfying $\int d^2x m({\vec x}) = 0$.  
The probability of an extreme fluctuation
in the system dipole moment may be expressed as
\begin{eqnarray}
{\cal F} \propto \sum_{\lbrace m \rbrace} 
e^{-H_{CG}[m]/T } \delta[Im~m({\vec k}_d) - \pi] \nonumber\\
=\int_{-\infty}^{\infty} d\lambda e^{-i\lambda \pi} 
\sum_{\lbrace m \rbrace} e^{-H_{CG}[m]/T } e^{i\lambda Im~m({\vec k}_d)} ,
\label{bvav}\\
\nonumber
\end{eqnarray}
where we have adopted the same definition of $P^x_{eff}$
as in Eq.\ref{peff}, with $\rho({\vec x})$ replaced by $m({\vec x})$,
and ${\vec k}_d \equiv (2\pi/L,0)$.
In the bound state
the integer nature of $m({\vec x})$ may not be ignored; however,
we can make progress by adopting the dual description of
the partition function \cite{jose}.  This involves employing the
Poisson resummation formula to rewrite
Eq. \ref{bvav} as 
\begin{equation}
{\cal F} \propto
\int_{-\infty}^{\infty} d\lambda e^{-i\lambda \pi}
\int {\cal D} \phi \sum_{\lbrace n \rbrace} 
e^{-H_{CG}[\phi]/T } e^{i\lambda Im~\phi({\vec k}_d)} 
e^{-2\pi i \sum_{\vec x} \phi({\vec x}) n({\vec x}) }.
\end{equation}
The integration over the continuous field $\phi$ may be 
carried through, with the result
\begin{equation}
{\cal F} \propto 
\int_{-\infty}^{\infty} d\lambda e^{-i\lambda \pi}
\sum_{\lbrace n \rbrace} 
e^{-{T \over {2K}} \sum_{\vec k} |2\pi n({\vec k}) - \lambda Im~n({\vec k}_d)
~\delta_{{\vec k},{\vec k}_d}|^2/ G(k)}.
\label{sos}
\end{equation}
A bound vortex fixed point is generated by replacing $K$ with a
renormalized value, and exchanging the
sum over integers $n$ in Eq. \ref{sos} with a functional
integral over continuous fields \cite{chaikin}.  The resulting
model represents the rough phase of a solid-on-solid model.
Once the integer sum has been replaced by an integral,
the $\lambda Im~n({\vec k}_d)$ term in the integrand may
be shifted away and the functional integral in fact has no
dependence on $\lambda$.  It then immediately follows
that ${\cal F}=0$ for the bound vortex phase.

These considerations lead us to expect that one should observe
large vortex dipole fluctuations in the unbound phase, but not in
the bound one.  We now demonstrate this is indeed the case
using a Langevin dynamics simulation.

{\it Simulation --}
Our simulations focus on the $XY$ model
for which we assign dynamics to the spins and coupling to
a heat bath to generate a distribution of configurations.
The equations of motion for our system are taken to be
\begin{equation}
\Gamma {{d^2 \theta({\vec x}) } \over {dt^2}} = 
{{\delta H_{XY}} \over {\delta \theta({\vec x}) } }
+ \zeta({\vec x}) - \eta {{d \theta({\vec x}) } \over {dt}}.
\label{eqmot}
\end{equation}
$\Gamma$ is an effective moment of inertia for the
$XY$ spins, which for simplicity we set to 1 in the simulations,
$\zeta$ is a random torque which models coupling to a
heat bath, and $\eta$ is a viscosity.   
To satisfy the fluctuation-dissipation theorem, the
random torques are drawn from a distribution satisfying
$<\zeta(\vec{r},t)  \zeta(\vec{r}',t')> = 2\eta T \delta_{\vec{r},\vec{r}'} \delta (t-t')$
with $T$ the temperature of the system.
Finally, our $XY$ Hamiltonian
is 
\begin{equation}
H_{XY} = -K \sum_{< {\vec x},{\vec x}'>} cos[\theta({\vec x}) - \theta({\vec x}') ]
- h \sum_{\vec x} cos \theta({\vec x})
\end{equation}
where we take the angles $\theta({\vec x})$ to reside on an
$L \times L$ square lattice.
To perform the simulation, we have discretized the
time derivative in Eq. \ref{eqmot} and used a standard
random number generator \cite{press} to generate
a realization of $\zeta(\vec{r},t)$ at each time step.
A typical run consists of $10^6$ Langevin sweeps
for equilibration, followed by $9\times 10^6$ measurement
steps.  In accumulating the data, we repeated 
runs for each set of parameters with $\sim10$ different
seeds, allowing us to estimate our statistical errors.
Simulations were performed for system sizes
as large as $L=199$, although most of the simulations
were in the range $19 \le L \le 59$.

Our measurement consists of counting the number
of times a component of the system dipole moment ${\vec P}_{ext}$ passes through
$(n  + {1 \over 2})L$ for $n$ any integer.
We then plot the number of such events divided
by the total simulation time, yielding a measure of the
fraction of configurations $\cal F$ for which the system has
attained its maximal value.
One advantage of the Langevin dynamics approach
is that the vortex dipole moment ${\vec P}_{ext}$ changes by several
steps with each Langevin sweep, but these steps are
always much smaller than $L$ except for very small
values of $L$.  This allows us to detect when ${\vec P}_{ext}$ has
passed through $(n  + {1 \over 2})L$ even if in the immediate time
step before and the step after ${\vec P}_{ext}$ was not measured to be
precisely at this value.  A larger
number of events can then be accumulated than one might
in a Monte Carlo simulation employing a cluster algorithm,
since the configurations generated in the latter are not
related in any simple way, forcing one to count only
configurations for which ${\vec P}_{ext}$ is precisely 
$(n  + {1 \over 2})L$.
Note that we count passages in both directions; this tells
us how often ${\vec P}_{red}$ visits its extremal value.

As a check on the method, we first present results of
simulations in the absence of the symmetry-breaking
field, for which vortices unbind in a Kosterlitz-Thouless transition.
Fig. 1 illustrates these for $L = 59$,
$h=0$ and $K=1$.
One can see $\cal F$ decreases
sharply as the temperature approaches $T \approx 0.9$
from above, so that $\cal F$ appears to be vanishing quite
close to the accepted value of $T^{XY}_{KT}=0.892$.  
The differing behavior of $\cal F$ above and below the transition
can be further confirmed by examining its size dependence:
for $T$ below the transition temperature, $\cal F$ decreases
with system size, apparently approaching zero as $L \rightarrow \infty$,
whereas for higher values of $T$, $\cal F$ {\it increases}, approaching
a constant value from below.  These differing size dependences
strongly support the idea that $\cal F$ distinguishes the bound and
unbound phases.

We now turn to the case $h>0$, for which we show some typical
results in  Fig. 2.  The behavior of $\cal F$ as a function of system size
takes three differing forms, depending on the values of $h$ and $T$.
At high temperature and low fields, $\cal F$ clearly approaches 
a non-vanishing value in the large system size limit (e.g., $h=0.18$ curve
in Fig. 2).  Unlike the $h=0$ case the asymptotic value of 
$\cal F$ is approached from above, indicating the that symmetry breaking
field has had some effect, although for small enough $h$ the
vortices remain unbound.  At intermediate values of $h$, we
find $\cal F$ decreasing as a power law in $1/L$ ($h=0.3$ curve),
indicating a bound vortex phase behaving as one would expect
for logarithmically interacting vortices.  Finally, at the largest
values of $h$ ($h=0.5$ curve), $\cal F$ decreases {\it exponentially}
with $L$, as might be expected for vortices interacting with a 
linear binding potential.

Our results indicate that in the presence of a magnetic field,
there exists an unbound vortex state and two different bound
vortex states in the $XY$ model.  This behavior is precisely
what was predicted in the $RG$ analysis of Ref. \cite{decon}.
At low temperatures, vortex-antivortex pairs are connected
by a string of overturned spins, leading to linear confinement.
As temperature is increased, fluctuations can lead to a 
roughening of the strings binding the vortices, driving the
effective string tension to zero.  The vortex-antivortex
pairs nevertheless retain a logarithmic attraction and
remain bound as the system passes through the transition.
At still higher temperatures, a second transition occurs in
which the vortices do ultimately unbind.  A remarkable
feature of these transitions is that they are extremely
continuous\cite{decon}, so that no singularity is expected
in the free energy as the transitions occur.  Measurements
of the specific heat and magnetization in our simulations
show no indications of any such singularities.  Thus, one
is forced to probe the vortices directly in order to 
detect their unbinding, as we have done by probing the
system vortex dipole moment.  It is interesting to 
note that, in the absence of singularities in the free-energy,
it is unclear whether vortex unbinding should be thought
of as a true thermodynamic phase transition.  To our knowledge, this
is the first example of a defect-unbinding transition
that does not have the full character of a phase transition.

In summary, we have developed a method of detecting vortex
unbinding in two-dimensional systems by tracking extreme
fluctuations in the vortex dipole moment of the system.
For a system of size linear $L$, the fraction of configurations $\cal F$
having a dipole moment component $P_{x}$ or  $P_{y}$
equal to its largest value consistent with periodic
boundary conditions ($L/2$) approaches a constant for
large sizes when vortices are unbound, and vanishes
when vortices are bound.  We have demonstrated this 
for the Kosterlitz-Thouless transition in $XY$ model,
for which $\cal F$ can locate the transition temperature
with reasonable accuracy.  In the presence of a
magnetic field tending to orient the spins, we have
found that there is an unbound vortex state and two
possible bound vortex states, one consistent with
logarithmic binding of the vortices, the other with
linear confinement.  The method presented here is
easily generalizable, and should be applicable to systems
with other topological, point-like defects.

{\it Acknowledgments --}
The authors are indebted to the
Center for Computational Sciences at the University of Kentucky
for providing computer time.  Helpful discussions with
Dr. Donald Priour are gratefully acknowledged.  This work was
supported by NSF Grant No. DMR-01-08451.

\newpage

\begin{figure}
\begin{center}
\includegraphics[scale=0.6]{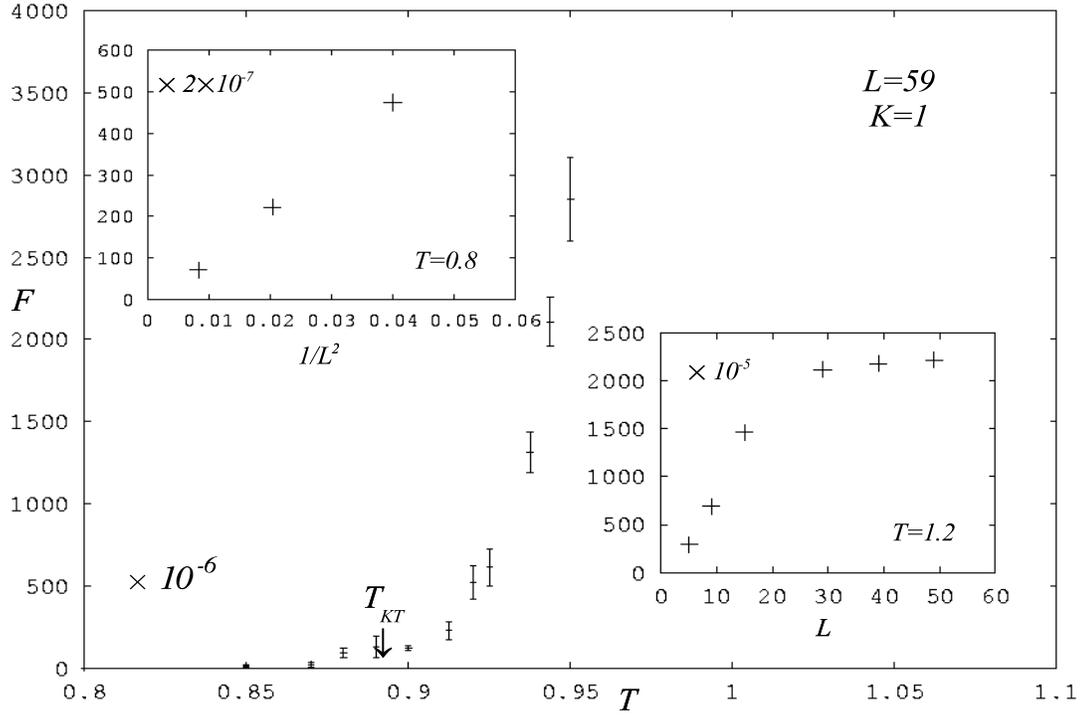}
\caption{
$F$ vs. temperature for $XY$ system with $h=0$, $L=59$.
$F$ drops sharply in vicinity of the known Kosterlitz-Thouless temperature.
Left inset: $F$ vs. $1/L^2$ for $T=0.8$, demonstrating $F$ vanishes
in the bound vortex state.  Right inset: $F$ vs. $L$ for $T=1.2$, demonstrating
$F$ approaches a non-zero constant in the unbound vortex state.
}
\end{center}
\end{figure}
%\vspace{-2.0in}
\newpage
\begin{figure}
\begin{center}
\includegraphics[scale=0.6,clip='true']{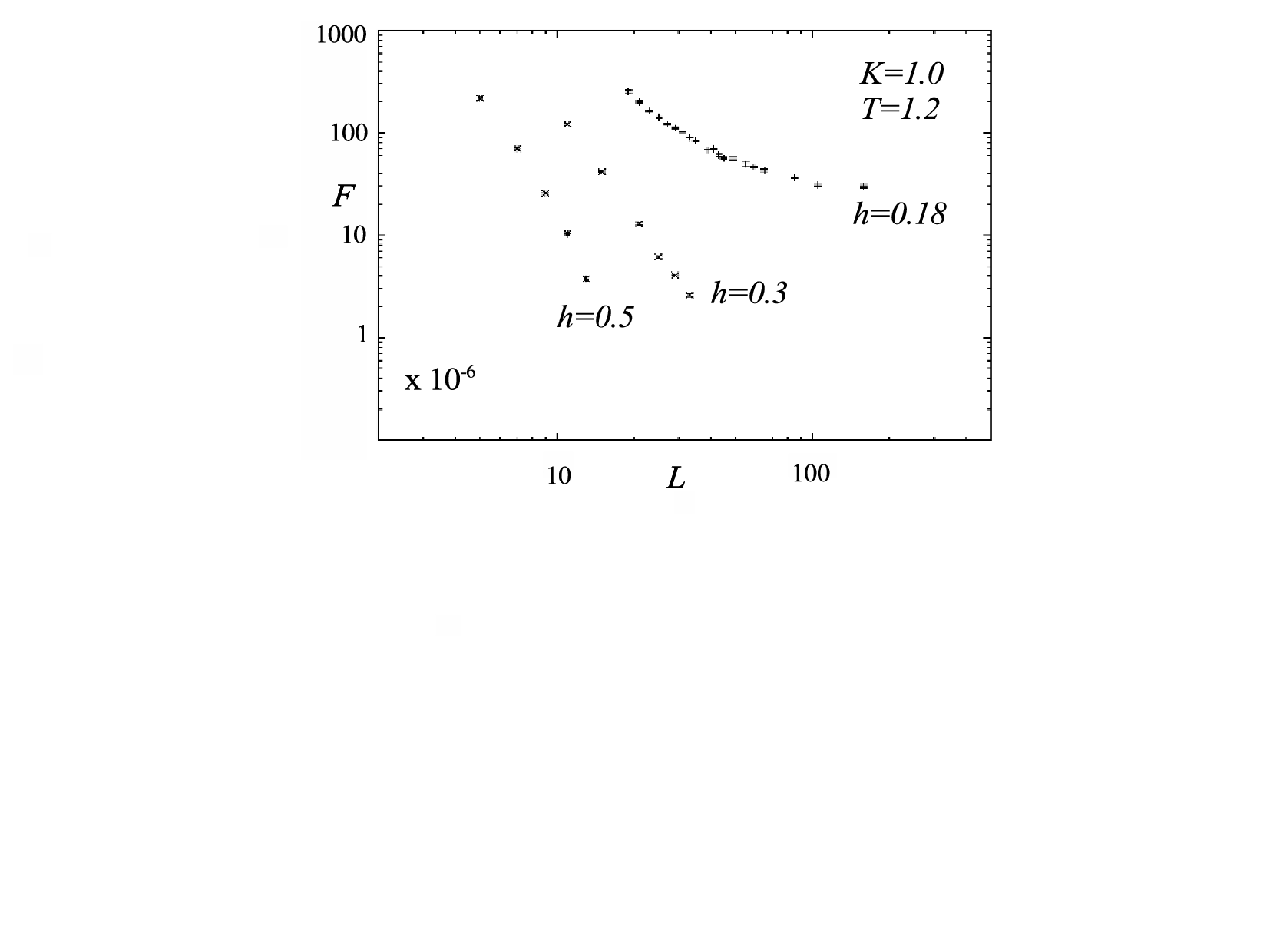}
\vspace{-2.0in}
\caption{
$F$ vs. system size for $XY$ system for $h=0.18,~0.30,~0.50$
for fixed temperature $T=1.2$  $F$ approaches a constant
for large system sizes for smallest field, vanishes as a power
law for intermediate field (straight line on log-log plot), and
vanishes exponentially for largest field (downward curvature
on log-log plot).  The three different behaviors indicate one
unbound vortex phase, logarithmically bound phase, and a
linearly confined phase.
}
\end{center}
\end{figure}

\begin{references}
\bibitem{huh?} The sum of the $\Delta \theta_{ij}$ does not
necessarily vanish.  This is a consequence of having constrained
these to the finite interval.
\end{references}
\end{document}